\documentclass[sigconf]{acmart}
\AtBeginDocument{%
  \providecommand\BibTeX{{%
    Bib\TeX}}}

\usepackage{enumitem}
\usepackage{graphicx}
\usepackage{subfigure}

\setcopyright{acmlicensed}
\copyrightyear{2025}
\acmYear{2025}
\acmDOI{10.1145/3726302.3731935}
\acmConference[SIGIR '25]{Proceedings of the 48th International ACM SIGIR Conference on Research and Development in Information Retrieval}{July 13--18, 2025}{Padua, Italy}
\acmBooktitle{Proceedings of the 48th International ACM SIGIR Conference on Research and Development in Information Retrieval (SIGIR '25), July 13--18, 2025, Padua, Italy}
\acmISBN{979-8-4007-1592-1/2025/07}

\begin{document}

\title{A Generative Re-ranking Model for List-level Multi-objective Optimization at Taobao}


\author{Yue Meng}
\email{mengyue.meng@taobao.com}
\author{Cheng Guo}
\email{mike.gc@taobao.com}
\affiliation{%
  \institution{Taobao \& Tmall Group of Alibaba}
  \city{Beijing}
  \country{China}
}

\author{Yi Cao}
\authornote{Corresponding author}
\email{dylan.cy@taobao.com}
\author{Tong Liu}
\email{yingmu@taobao.com}
\affiliation{%
  \institution{Taobao \& Tmall Group of Alibaba}
  \city{Hangzhou}
  \country{China}
}

\author{Bo Zheng}
\email{bozheng@alibaba-inc.com}
\affiliation{%
  \institution{Taobao \& Tmall Group of Alibaba}
  \city{Beijing}
  \country{China}
}

\renewcommand{\shortauthors}{Yue Meng, Cheng Guo, Yi Cao, Tong Liu, \& Bo Zheng}

\begin{abstract}
  E-commerce recommendation systems aim to generate ordered lists of items for customers, optimizing multiple business objectives, such as clicks, conversions and Gross Merchandise Volume (GMV). Traditional multi-objective optimization methods like formulas or Learning-to-rank (LTR) models take effect at item-level, neglecting dynamic user intent and contextual item interactions. List-level multi-objective optimization in the re-ranking stage can overcome this limitation, but most current re-ranking models focus more on accuracy improvement with context. In addition, re-ranking is faced with the challenges of time complexity and diversity. In light of this, we propose a novel end-to-end generative re-ranking model named Sequential Ordered Regression Transformer-Generator (SORT-Gen) for the less-studied list-level multi-objective optimization problem. Specifically, SORT-Gen is divided into two parts: 1)Sequential Ordered Regression Transformer innovatively uses Transformer and ordered regression to accurately estimate multi-objective values for variable-length sub-lists. 2)Mask-Driven Fast Generation Algorithm combines multi-objective candidate queues, efficient item selection and diversity mechanism into model inference, providing a fast online list generation method. Comprehensive online experiments demonstrate that SORT-Gen brings +4.13\% CLCK and +8.10\% GMV for Baiyibutie, a notable Mini-app of Taobao. Currently, SORT-Gen has been successfully deployed in multiple scenarios of Taobao App, serving for a vast number of users.
\end{abstract}





\maketitle

\section{Introduction}
In e-commerce recommendation systems, it is essential to optimize multiple business objectives, such as clicks, conversions and Gross Merchandise Volume (GMV). Current practices adopt two paradigms. One involves designing a manually formula (e.g. $CTR^a*CVR^b*price^c$), the other involves building a Learning-To-Rank (LTR) model that treat objectives as training features. Both merge multiple objectives into a single rank score based on weight parameters for greedy item sorting. Unreasonable weight settings negatively affect multiple objectives. To tackle this, various methods are proposed to identify the Pareto frontier and obtain Pareto optimal solutions\cite{DBLP:books/daglib/0021267}. For the proposed formulation, optimization theory can be employed to determine the optimal weight parameters in an offline manner. Alternatively, bandit algorithms or reinforcement learning methods are utilized for online weight parameter searching. For stacking a LTR model, methods like Online Deep Controllable LTR (DC-LTR)\cite{DBLP:conf/aaai/ZengYHNLZM21} and Pareto Efficient LTR (PE-LTR)\cite{DBLP:conf/recsys/LinCPSXSZOJ19} employ linear programming to dynamically adjust loss weights in conjunction with the ranking models\cite{DBLP:conf/icml/BurgesSRLDHH05}\cite{DBLP:journals/sigir/JoachimsLLZ07}. These approaches which employ fixed weight parameters centered on item-level multi-objective optimization, give rise to two critical issues.

First, dynamic user interests mismatch static objective weights. During the browsing process, user interests fluctuate dynamically within the session\cite{DBLP:conf/kdd/XuCWYSWHLZGH23}. For example, early positions may prioritize conversion, while later ones focus on exploratory clicks. Existing methods are limited in their ability to adapt to such granularity 

Second, item-level multi-objective optimization ignores contextual item interactions\cite{DBLP:conf/sigir/AiBGC18}. Current approaches maximize cumulative item-level utility but disregard psychological and economic principles (e.g., anchoring, contrast effects). In reality, users' decisions on one item depend on surrounding options.

In a multi-stage recommendation system, the final stage known as re-ranking stage\cite{DBLP:conf/ijcai/LiuXQS00ZT22}\cite{DBLP:conf/recsys/PeiZZSLSWJGOP19}, is applied to address the above two limitations. However, current re-ranking models like PRM\cite{DBLP:conf/recsys/PeiZZSLSWJGOP19}, Seq2Slate\cite{DBLP:journals/corr/abs-1810-02019}, DLCM\cite{DBLP:conf/sigir/AiBGC18} and GRN\cite{DBLP:journals/corr/abs-2104-00860} focus on leveraging contextual information to improve accuracy but neglect list-level multi-objective optimization, which is actually a relatively less-studied problem. To the best of our knowledge, industry has only made some simple rule-based attempts in the re-ranking stage: 1) Combinatorial Efficiency. Directly evaluating all permutations of candidate lists ( $O(n!)$complexity) is infeasible for latency-sensitive systems. A straightforward approach would involve repeatedly calling the context-aware model for greedy item selection. Such iterative model calls introduce unacceptable overhead. 2) Accuracy-Diversity Trade-off\cite{DBLP:conf/kdd/HuangWZX21}\cite{DBLP:conf/kdd/LinWMZWJW22}\cite{DBLP:conf/www/ChengWMSX17}. Over-optimizing multi-objective scores risks homogenizing the list, degrading user experience\cite{DBLP:conf/recsys/ZhangH08}.

Compared to ranking, re-ranking shifts the modeling of user lifecycle interest to a list-level real-time interest. Ranking models like Deep Interest Network (DIN)\cite{DBLP:conf/kdd/ZhouZSFZMYJLG18} and Search-based User Interest Model (SIM)\cite{DBLP:conf/cikm/PiZZWRFZG20} both use Attention mechanisms to encode behavior sequences. Recently, due to the popularity of Large Language Model (LLM), some generative models have directly employed Transformer to model user behavior sequences, such as Meta's generative recommendation\cite{DBLP:conf/icml/ZhaiLLWLCGGGHLS24}. Re-ranking models are naturally suited to this generative architecture.

To this end, we propose a novel generative re-ranking model named Sequential Ordered Regression Transformer-Generator (SORT-Gen) for joint list-level optimization in e-commerce feed recommendation, which consists of two main components. 1)Sequential Ordered Regression Transformer models real-time user intent via Transformer. It innovatively uses ordered regression to accurately predict multi-objective values for variable-length sub-lists. 2)Mask-Driven Fast Generation Algorithm provides an efficient online method for list generation. It eliminates iterative inference by integrating candidate cross-distribution and diversity into tensor operations. Mask matrices are deployed to guide item selection in one forward pass.



To sum up, the main contributions of our work as follows:
\begin{itemize}
\item We investigate less-studied list-level multi-objective optimization problem, which is significant in the industry.
\item We introduce SORT-Gen, an efficient generative re-ranking model, offering practical solutions for list-level multi-objective optimization on large-scale recommendation systems.
\item We conduct experiments on online A/B testing. Results prove that SORT-Gen brings huge benefits and has been deployed in multiple scenarios of Taobao App.
\end{itemize}

\section{Problem Formula}

A typical recommendation system usually consists of three main stages, i.e., matching, ranking and re-ranking\cite{DBLP:conf/cikm/WilhelmRBJCG18}. Denote the result from ranking stage as $I = \left\{i_1, i_2, ..., i_{l_s}\right\}$ with length $l_s$. Re-ranking aims to reorder $I$ so as to return an optimal ordered list $R^* = \left\{i^*_1, i^*_2, ..., i^*_{l_o}\right\}$, where $l_o$ refers to actual length of recommendation results. In practice, $l_s$ is always larger than $l_o$.

Introducing the context $C$, when $R'$ is a potential reordering result, $V_{click}$, $V_{conversion}$ and $V_{GMV}$ represent the measurement of click value, conversion value and GMV value at list-level respectively, we can formally express re-ranking that integrates multi-objective optimization as follows: 
\begin{equation}
R^* = argmax_{R'} [ \alpha * V_{click}(R', C) + \beta * V_{conversion}(R', C) + \gamma * V_{GMV}(R', C)] 
\end{equation}
where $\alpha$, $\beta$ and $\gamma$ represent the trade-offs at Pareto boundaries for commercial purposes. List value estimation with context is structured as a sequential incremental value accumulation approach. For example, $V_{click}$ is expressed as:
\begin{equation}
V_{click}(R', C) = v_{click}(i'_1) + v_{click}(i'_2 | [i'_1]) + ... + v_{click}(i'_{l_o} | [i'_1, ..., i'_{l_o-1}])
\end{equation}
where $v_{click}$ refers to item-level click-value at each step.

\section{The Proposed Model}

\begin{figure*}[htbp]
\centering
\includegraphics[width=\linewidth]{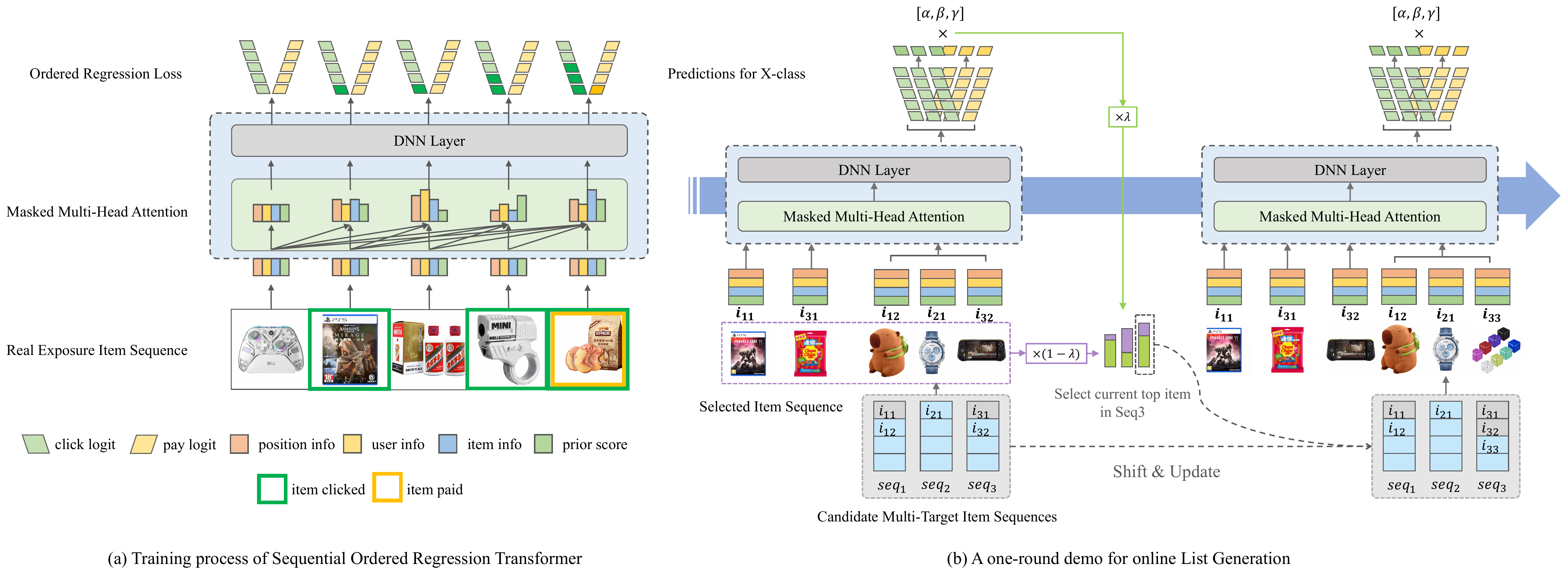}
\caption{An overview of the SORT-Gen framework.}
\label{fig1}
\end{figure*}

\subsection{Sequential Ordered Regression Transformer}

\subsubsection{Model Structure}
As illustrated in Fig.1(a), the SORT method employs real exposure item lists as training samples. Four categories of features are employed: $Item\ Info$, $User\ Info$, $Position\ Info$ and $Prior\ Score$. $Item\ Info$ is organized in a sequential format. A raw user feature vector $E'_{user} \in \mathrm{R}^{n \times d_{user}}$ is replicated $l_o$ times to form a temporal feature matrix $E_{user} \in \mathrm{R}^{n \times l_o \times d_{user}}$. In order to leverage the sequential information, we inject $Position\ Info$ as a learnable position embedding $E_{position} \in \mathrm{R}^{n \times l_o \times d_{position}}$ into the input embedding. Estimated CTR and CVR denoted as $E_{score} \in \mathrm{R}^{n \times l_o \times d_{score}}$ are derived from previous ranking models\cite{DBLP:conf/www/XiaCHLL23}\cite{DBLP:conf/sigir/HouZCL24}, encapsulates rich information about user-item interactions and historical user behaviors. Input vector can be formalized as:
\begin{equation}
E_{input} = E_{item} \oplus E_{position} \oplus E_{user} \oplus E_{score} \in \mathrm{R}^{n \times l_o \times d}
\end{equation}
where $\oplus$ is the concatenate operator.


SORT-Gen takes Transformer as main architecture\cite{DBLP:conf/nips/VaswaniSPUJGKP17}. Pre-Norm residual connection, applying Layer Normalization before each block, is adopted for better training stability\cite{DBLP:conf/iwslt/NguyenS19}\cite{DBLP:conf/acl/WangLXZLWC19}. To eliminate the influence of subsequent items, causal mask is employed\cite{DBLP:conf/cvpr/YangZQ021}.

Temporal input matrix $E_{input} \in \mathrm{R}^{n \times l_o \times d}$ is projected into the transformer's hidden dimension. A standard transformer architecture consisting of Feed-Forward Network (FFN) and Multi-Head Self-Attention (MHSA) with causal mask follows. Output at each position is then passed through two separate Multi-Layer Perceptrons (MLPs) for predicting click and conversion objectives respectively.

\subsubsection{Ordered Regression Loss}

We found that ordered regression is an effective loss function for step-by-step list generation, accounting for the sequential incremental values when empirically evaluating sub-lists of increasing length across multiple rounds. 

For each pair of adjacent targets, a cross-entropy loss function is constructed, resulting in $l_o$ binary classification tasks for $l_o + 1$ categories. The overall loss function $
L(\theta) $ is represented as the sum of all effective cross-entropy loss functions:
\begin{equation}
L(\theta)=\sum_{i=1}^{l} \sum_{j=1}^l L_{i, j}^{C L I C K}\left(\theta_{i, j}\right)+\sum_{i=1}^{l} \sum_{j=1}^l L_{i, j}^{P A Y}\left(\theta_{i, j}\right)
\end{equation}
Here, $L_{i, j}$ is the loss function for the sub-list of length $j$ to judge whether it has reached $i$ clicks or conversions:
\begin{equation}
L_{i, j}\left(\theta_j\right)=-\sum_{k=1}^N\left[y_k<i\right] \log \left(1-p_{i, j}\left(x_k\right)\right)+\left[y_k \geq i\right] \log \left(p_{i, j}\left(x_k\right)\right)
\end{equation}
Where N is the number of samples, $y_k$ denote the true count of corresponding actions, and $p_{i, j}\left(x_k\right)=\sigma\left(x_k^T \theta_{i, j}\right)$ represents the predicted probability that the $k$-th sample's corresponding actions exceeding $i$ in the sub-list of length $j$. The term $\left[condition\right]$ is an indicator function that equals 1 if the condition holds, and 0 otherwise.The incremental value is expressed as $p_{i, j}\left(x_k\right)-p_{i+1, j}\left(x_k\right)$.

\subsection{Mask-Driven Fast Generation Algorithm}

To meet Taobao’s strict latency requirements, we design the inference pipeline shown in Fig. 1(b). Traditional item selection (invoking SORT $l_o$ times with $l_s$ candidates) incurs prohibitive network overhead, prolonging response time(RT) by $l_o$ times. Our Mask-Driven Fast Generation Algorithm not only eliminates iterative calls by item selection via tensor operations, reducing RT to near ranking-stage level, but also enables multi-objective cross-distribution.

\subsubsection{Multi-objective Candidate Queues}

Initially, we partition the candidate pool $I = \left\{i_1, i_2, ..., i_{l_s}\right\}$ into multiple distinct objective-specific ordered queues. For example, we utilized three separate queues for different objectives: click, conversion and GMV. Additionally, it is feasible to create composite objective queues, such as $0.5 * CTR + 0.5 * CTR * CVR$, etc. Each queue ranks items by predefined scores. During list generation, we iteratively select top items from these queues, reducing the search space from $l_s$ to the actual number of queues. The maximum size of each queue is set to $l_o$ to satisfy the requirements of the generation process.

Naturally, there are two strategies for partition: Depth-First Search (DFS) and Breadth-First Search (BFS). DFS prioritizes the filling of queues associated with high-priority objectives, whereas BFS prioritizes the highest-ranked items. In practice, we have found that when differentiation is sufficiently large, both methods converge to equivalent allocations.

\subsubsection{Efficient Item Selection}
To eliminate iterative model invocations, Mask-Driven Fast Generation Algorithm simulates multi-round candidate selection within one forward pass via:
\begin{enumerate}[label=\roman*)]
\item Monitor selected items and queue states by mask matrices.
\item Expand sub-lists by concatenating current selections with top candidates from each queue, then compute $V(R', C)$ in mini-batches.
\item Choose the candidate maximizing $V(R', C)$, and shift the queue by updating masks for next iteration.
\end{enumerate}

As shown in Fig. 1(b), all $l_o$ steps are compressed into a single call using tensor operations, ultimately generating ordered results.


\subsubsection{Efficiency-Diversity Equilibrium}
Purely value-driven list generation poses risks of homogenizing recommendations, leading to an accuracy-diversity dilemma\cite{DBLP:conf/kdd/LinWMZWJW22}.

To avoid post-processing, we integrate Maximal Marginal Relevance (MMR)\cite{DBLP:conf/airs/YangJL05} directly into Mask-Driven Fast Generation Algorithm, allowing for joint optimization of accuracy and diversity:
\begin{equation}
    \mathrm{MMR} = \arg \max _{i_a \in Q \backslash S}\left[\lambda * \operatorname{V}\left([S,i_a], S\right)-(1-\lambda)*\max_{i_b \in S}\operatorname{SIM}\left(i_a, i_b\right)\right]
\end{equation}
Where $S$ represents currently selected sub-list, $Q$ is the current set of candidate items from the top of queues and $\max_{i_b \in S}\operatorname{SIM}\left(i_a, i_b\right)$ is the maximum cosine similarity between candidate items and selected items via pre-trained multi-modal embeddings. During item selection, we penalize candidates overly similar to existing sub-list within a sliding window (simulating user's browsing perspective). When expanding the sub-list, item selection criterion is modified from $V(R', C)$ to $MMR$. 

\section{Experiments}

\subsection{Online Experiment Results}
Compared to traditional re-ranking models, SORT-Gen implements list-level multi-objective optimization targeting objective permutations, which are unattainable through standard offline metrics(e.g., AUC, NDCG). Its efficacy is validated via online A/B tests on Taobao’s massive live traffic, convincingly proving its superiority.
\subsubsection{Experiment Setup}

\begin{table}
  \caption{Online experiment results. All values are the relative improvements over greedy selection based on the formula.}
  \label{tab:freq}
  \begin{tabular}{p{4cm}ccc}
    \toprule
    Method & CLICK & ORDER & GMV\\
    \midrule
    LTR & -0.06\% & +2.07\% & +0.60\%\\
    Pareto Efficient Formulas & -0.19\% & -0.16\% & +3.47\%\\
    \hline
    fastDPP & +3.28\% & +2.27\% & +0.88\%\\
    PRM & +2.63\% & +2.21\% & +0.17\% \\
    FFT Context-aware Model & +3.43\% & +4.56\% & 2.48\% \\
    FFT Context-aware Model + fastDPP & +5.26\% & +8.31\% & +5.15\% \\
    \textbf{SORT-Gen} & \textbf{+9.61\%*} & \textbf{+8.35\%*} & \textbf{+13.67\%*}\\
  \bottomrule
\end{tabular}
\end{table}

We compare two item-level multi-objective methods, which can actually coexist with SORT-Gen, and four effective re-ranking methods. Our full-deployed baseline re-ranking model is an advanced method that stood out as the optimal model in practice before SORT-Gen was proposed. It initially models the context sequence using Fourier Transform\cite{DBLP:conf/www/ZhouYZW22} and Attention Mechanism to construct a FFT Context-aware CTR Prediction Model. This is also regarded as an improved version of PRM\cite{DBLP:conf/recsys/PeiZZSLSWJGOP19}, which is widely considered as a baseline for re-ranking. Then, it employs fastDPP\cite{DBLP:conf/nips/ChenZZ18} to generate a diversified item list. 

We set $\alpha$, $\beta$ and $\gamma$ to 5, 1, 1 respectively in the experiment. They only work in the inference phase so that they need to be manually selected and adjusted online based on scene characteristics and business goals. Training weights of different objectives are not sensitive within a small range. We set them all to 1.

\subsubsection{Experiment Results}

Online A/B experiments have been launched on Baiyibutie(a notable mini-app of Taobao) over two weeks. Results are recorded in Table 1. SORT-Gen performs significantly better than other methods, as evidenced by the substantial improvement in multiple objectives. Compared to FFT Context-aware Model + fastDPP, SORT-Gen brings +4.13\% CLICK and +8.10\% GMV.


\subsubsection{Online Efficiency}
With the optimization introduced in 3.2, SORT-Gen's end-to-end latency reaches 19 ms—on par with baseline ranking systems and within industrial latency budgets.

\subsection{Ablation Study}

Table 2 quantifies the effects of different modules. All values are the relative improvements of complete SORT-Gen compared with modified SORT-Gen whose partial module is replaced. "Template method" refers to hard-coded rules multiple-objective cross-distribution without utilizing a learnable model. "Ranking top queue" involves changing the candidates to items solely by point-wise ranking scores. "Point-wise classification" means optimizing per-position cross-entropy loss independently. ESMMs mask the click loss for partially completed conversions. Without integrated MMR or other diversity mechanisms, there are some cases that homogeneous items are exposed at the same time, which would hurt the user experience and multi-objective values to some extent.

\begin{table}
  \caption{Results of ablation study.}
  \label{tab:freq}
  \begin{tabular}{p{4.5cm}ccc}
    \toprule
    Compared Method & CLICK & ORDER & GMV\\
    \midrule
    vs. Multi-objective queues + template method & +10.39\% & -0.30\% & +2.85\%\\
    vs. SORT-Gen w/o multi-objective queue + ranking top queue  & +8.30\% & +1.81\% & +2.15\%\\
    vs. SORT-Gen w/o ordered regression + point-wise classification  & +5.06\% & +1.60\% & +5.11\%\\
    vs. SORT-Gen w/o ordered regression + ESMM-like modeling loss  & +8.38\% & +2.34\% & +2.34\%\\
    vs. SORT-Gen w/o integrated MMR & +3.82\% & -0.84\% & +4.28\%\\
  \bottomrule
\end{tabular}
\end{table}

\subsection{Data Analysis}

\begin{figure}[h]
  \centering
  \includegraphics[width=\linewidth]{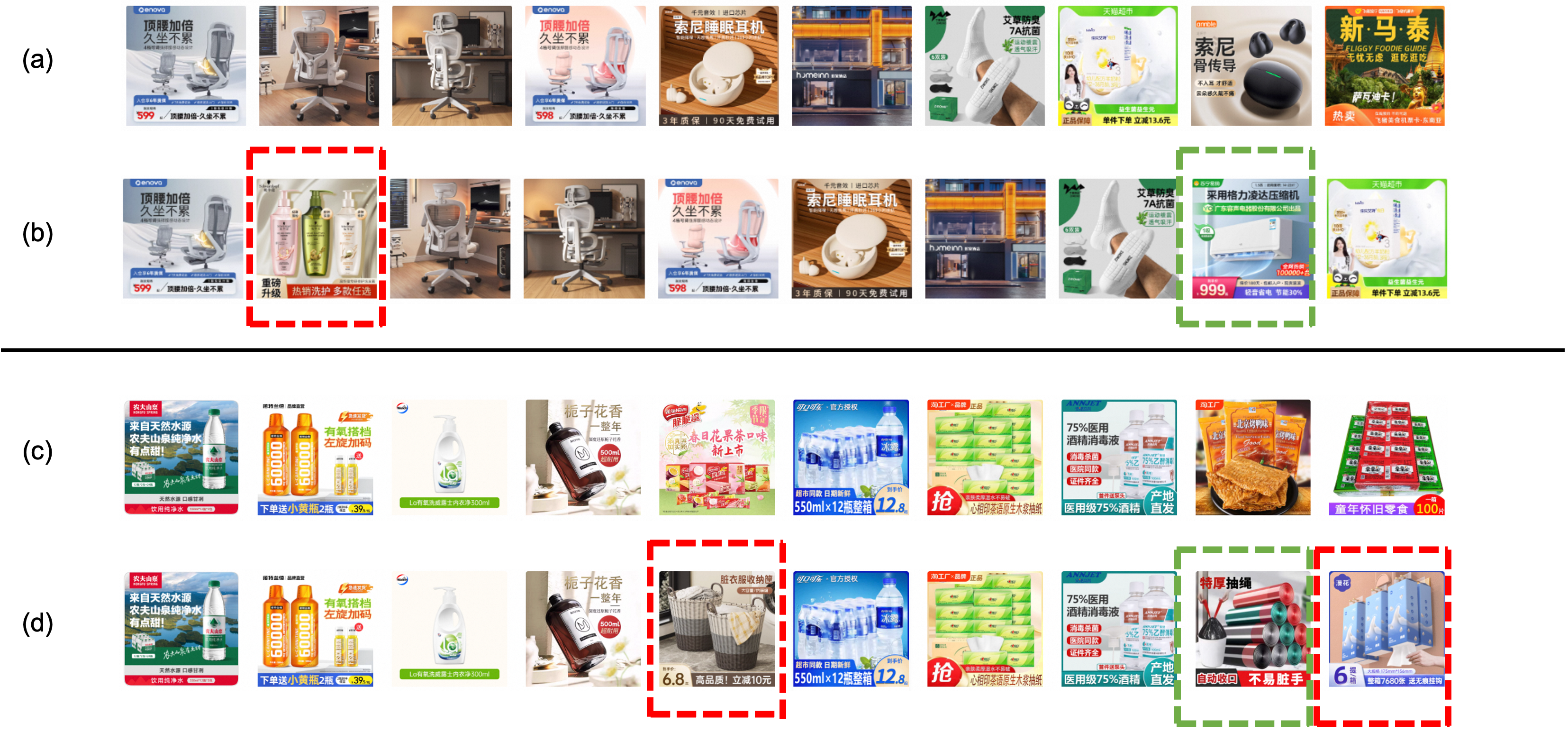}
  \caption{Online real-world cases where SORT-Gen works.}
\end{figure}
Fig. 2 illustrates two specific cases of SORT-Gen. (a) and (b) show a result with the index $[0, 1, 0, 0, 0, 0, 0, 0, 2, 0]$, where $0, 1, 2$ represents different multi-objective candidate queues. (a) is the item list prioritized by the rank scores. (b) is the final item list generated by SORT-Gen, with the red box and green box representing items from other two candidate queues. Similarly, (c) and (d) represent another case with the index $[0, 0, 0, 0, 1, 0, 0, 0, 2, 1]$.

\begin{figure}[h]
  \centering
  \includegraphics[width=\linewidth]{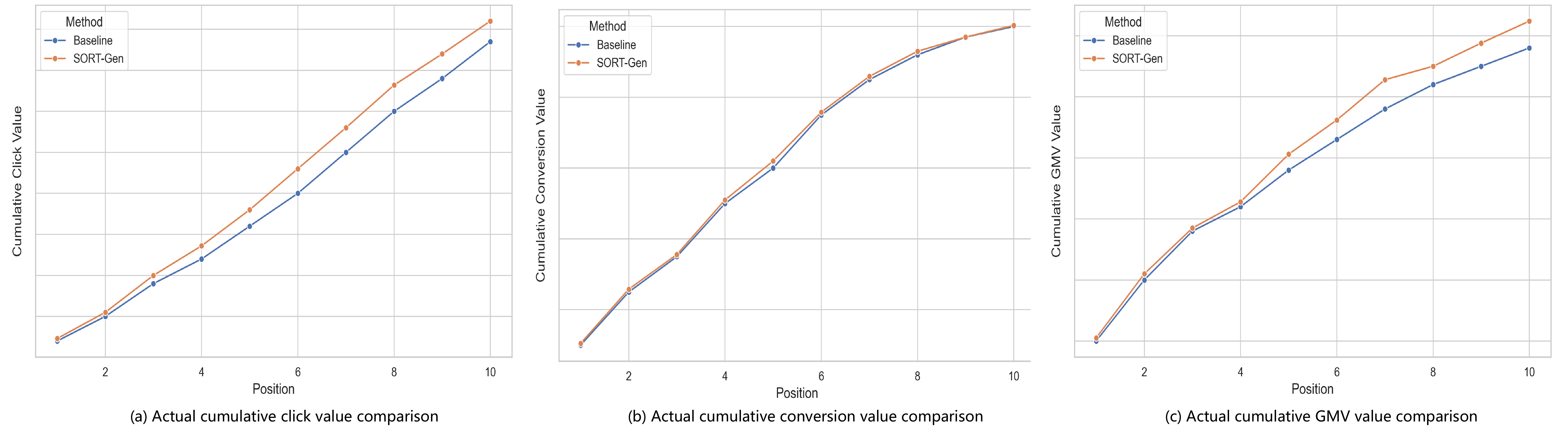}
  \caption{Multi-objective value based on real logs.}
\end{figure}


In addition, we collected and analyzed all online logs during the A/B online experiment to obtain the cumulative multi-objective value of sub-lists of varying lengths where specific values are anonymous. Results are shown in Fig. 3. As position increases, the gap of click value and GMV value between SORT-Gen and the baseline gradually increase, while the conversion value remains close.

\section{Conclusion}
In this paper, we identify a significant yet under-explored challenge in recommendation systems: list-level multi-objective optimization problem. To address this, We propose a novel generative re-ranking framework called SORT-Gen. Online A/B experiments conducted in real-world industrial e-commerce recommendation system demonstrate its superiority. Currently, SORT-Gen has been deployed in various scenarios of Taobao App, serving a vast number of users.

\bibliographystyle{ACM-Reference-Format}
\bibliography{sample-base}


\begin{thebibliography}{28}


\ifx \showCODEN    \undefined \def \showCODEN     #1{\unskip}     \fi
\ifx \showISBNx    \undefined \def \showISBNx     #1{\unskip}     \fi
\ifx \showISBNxiii \undefined \def \showISBNxiii  #1{\unskip}     \fi
\ifx \showISSN     \undefined \def \showISSN      #1{\unskip}     \fi
\ifx \showLCCN     \undefined \def \showLCCN      #1{\unskip}     \fi
\ifx \shownote     \undefined \def \shownote      #1{#1}          \fi
\ifx \showarticletitle \undefined \def \showarticletitle #1{#1}   \fi
\ifx \showURL      \undefined \def \showURL       {\relax}        \fi
\providecommand\bibfield[2]{#2}
\providecommand\bibinfo[2]{#2}
\providecommand\natexlab[1]{#1}
\providecommand\showeprint[2][]{arXiv:#2}

\bibitem[Ai et~al\mbox{.}(2018)]%
        {DBLP:conf/sigir/AiBGC18}
\bibfield{author}{\bibinfo{person}{Qingyao Ai}, \bibinfo{person}{Keping Bi}, \bibinfo{person}{Jiafeng Guo}, {and} \bibinfo{person}{W.~Bruce Croft}.} \bibinfo{year}{2018}\natexlab{}.
\newblock \showarticletitle{Learning a Deep Listwise Context Model for Ranking Refinement}. In \bibinfo{booktitle}{\emph{The 41st International {ACM} {SIGIR} Conference on Research {\&} Development in Information Retrieval, {SIGIR} 2018, Ann Arbor, MI, USA, July 08-12, 2018}}, \bibfield{editor}{\bibinfo{person}{Kevyn Collins{-}Thompson}, \bibinfo{person}{Qiaozhu Mei}, \bibinfo{person}{Brian~D. Davison}, \bibinfo{person}{Yiqun Liu}, {and} \bibinfo{person}{Emine Yilmaz}} (Eds.). \bibinfo{publisher}{{ACM}}, \bibinfo{pages}{135--144}.
\newblock
\href{https://doi.org/10.1145/3209978.3209985}{doi:\nolinkurl{10.1145/3209978.3209985}}


\bibitem[Bello et~al\mbox{.}(2018)]%
        {DBLP:journals/corr/abs-1810-02019}
\bibfield{author}{\bibinfo{person}{Irwan Bello}, \bibinfo{person}{Sayali Kulkarni}, \bibinfo{person}{Sagar Jain}, \bibinfo{person}{Craig Boutilier}, \bibinfo{person}{Ed~Huai{-}hsin Chi}, \bibinfo{person}{Elad Eban}, \bibinfo{person}{Xiyang Luo}, \bibinfo{person}{Alan Mackey}, {and} \bibinfo{person}{Ofer Meshi}.} \bibinfo{year}{2018}\natexlab{}.
\newblock \showarticletitle{Seq2Slate: Re-ranking and Slate Optimization with RNNs}.
\newblock \bibinfo{journal}{\emph{CoRR}}  \bibinfo{volume}{abs/1810.02019} (\bibinfo{year}{2018}).
\newblock
\showeprint[arXiv]{1810.02019}
\urldef\tempurl%
\url{http://arxiv.org/abs/1810.02019}
\showURL{%
\tempurl}


\bibitem[Burges et~al\mbox{.}(2005)]%
        {DBLP:conf/icml/BurgesSRLDHH05}
\bibfield{author}{\bibinfo{person}{Christopher J.~C. Burges}, \bibinfo{person}{Tal Shaked}, \bibinfo{person}{Erin Renshaw}, \bibinfo{person}{Ari Lazier}, \bibinfo{person}{Matt Deeds}, \bibinfo{person}{Nicole Hamilton}, {and} \bibinfo{person}{Gregory~N. Hullender}.} \bibinfo{year}{2005}\natexlab{}.
\newblock \showarticletitle{Learning to rank using gradient descent}. In \bibinfo{booktitle}{\emph{Machine Learning, Proceedings of the Twenty-Second International Conference {(ICML} 2005), Bonn, Germany, August 7-11, 2005}} \emph{(\bibinfo{series}{{ACM} International Conference Proceeding Series}, Vol.~\bibinfo{volume}{119})}, \bibfield{editor}{\bibinfo{person}{Luc~De Raedt} {and} \bibinfo{person}{Stefan Wrobel}} (Eds.). \bibinfo{publisher}{{ACM}}, \bibinfo{pages}{89--96}.
\newblock
\href{https://doi.org/10.1145/1102351.1102363}{doi:\nolinkurl{10.1145/1102351.1102363}}


\bibitem[Chen et~al\mbox{.}(2018)]%
        {DBLP:conf/nips/ChenZZ18}
\bibfield{author}{\bibinfo{person}{Laming Chen}, \bibinfo{person}{Guoxin Zhang}, {and} \bibinfo{person}{Eric Zhou}.} \bibinfo{year}{2018}\natexlab{}.
\newblock \showarticletitle{Fast Greedy {MAP} Inference for Determinantal Point Process to Improve Recommendation Diversity}. In \bibinfo{booktitle}{\emph{Advances in Neural Information Processing Systems 31: Annual Conference on Neural Information Processing Systems 2018, NeurIPS 2018, December 3-8, 2018, Montr{\'{e}}al, Canada}}, \bibfield{editor}{\bibinfo{person}{Samy Bengio}, \bibinfo{person}{Hanna~M. Wallach}, \bibinfo{person}{Hugo Larochelle}, \bibinfo{person}{Kristen Grauman}, \bibinfo{person}{Nicol{\`{o}} Cesa{-}Bianchi}, {and} \bibinfo{person}{Roman Garnett}} (Eds.). \bibinfo{pages}{5627--5638}.
\newblock
\urldef\tempurl%
\url{https://proceedings.neurips.cc/paper/2018/hash/dbbf603ff0e99629dda5d75b6f75f966-Abstract.html}
\showURL{%
\tempurl}


\bibitem[Cheng et~al\mbox{.}(2017)]%
        {DBLP:conf/www/ChengWMSX17}
\bibfield{author}{\bibinfo{person}{Peizhe Cheng}, \bibinfo{person}{Shuaiqiang Wang}, \bibinfo{person}{Jun Ma}, \bibinfo{person}{Jiankai Sun}, {and} \bibinfo{person}{Hui Xiong}.} \bibinfo{year}{2017}\natexlab{}.
\newblock \showarticletitle{Learning to Recommend Accurate and Diverse Items}. In \bibinfo{booktitle}{\emph{Proceedings of the 26th International Conference on World Wide Web, {WWW} 2017, Perth, Australia, April 3-7, 2017}}, \bibfield{editor}{\bibinfo{person}{Rick Barrett}, \bibinfo{person}{Rick Cummings}, \bibinfo{person}{Eugene Agichtein}, {and} \bibinfo{person}{Evgeniy Gabrilovich}} (Eds.). \bibinfo{publisher}{{ACM}}, \bibinfo{pages}{183--192}.
\newblock
\href{https://doi.org/10.1145/3038912.3052585}{doi:\nolinkurl{10.1145/3038912.3052585}}


\bibitem[Feng et~al\mbox{.}(2021)]%
        {DBLP:journals/corr/abs-2104-00860}
\bibfield{author}{\bibinfo{person}{Yufei Feng}, \bibinfo{person}{Binbin Hu}, \bibinfo{person}{Yu Gong}, \bibinfo{person}{Fei Sun}, \bibinfo{person}{Qingwen Liu}, {and} \bibinfo{person}{Wenwu Ou}.} \bibinfo{year}{2021}\natexlab{}.
\newblock \showarticletitle{{GRN:} Generative Rerank Network for Context-wise Recommendation}.
\newblock \bibinfo{journal}{\emph{CoRR}}  \bibinfo{volume}{abs/2104.00860} (\bibinfo{year}{2021}).
\newblock
\showeprint[arXiv]{2104.00860}
\urldef\tempurl%
\url{https://arxiv.org/abs/2104.00860}
\showURL{%
\tempurl}


\bibitem[Hou et~al\mbox{.}(2024)]%
        {DBLP:conf/sigir/HouZCL24}
\bibfield{author}{\bibinfo{person}{Chaoqun Hou}, \bibinfo{person}{Yuanhang Zhou}, \bibinfo{person}{Yi Cao}, {and} \bibinfo{person}{Tong Liu}.} \bibinfo{year}{2024}\natexlab{}.
\newblock \showarticletitle{{ECAT:} {A} Entire space Continual and Adaptive Transfer Learning Framework for Cross-Domain Recommendation}. In \bibinfo{booktitle}{\emph{Proceedings of the 47th International {ACM} {SIGIR} Conference on Research and Development in Information Retrieval, {SIGIR} 2024, Washington DC, USA, July 14-18, 2024}}, \bibfield{editor}{\bibinfo{person}{Grace~Hui Yang}, \bibinfo{person}{Hongning Wang}, \bibinfo{person}{Sam Han}, \bibinfo{person}{Claudia Hauff}, \bibinfo{person}{Guido Zuccon}, {and} \bibinfo{person}{Yi~Zhang}} (Eds.). \bibinfo{publisher}{{ACM}}, \bibinfo{pages}{2885--2889}.
\newblock
\href{https://doi.org/10.1145/3626772.3661348}{doi:\nolinkurl{10.1145/3626772.3661348}}


\bibitem[Huang et~al\mbox{.}(2021)]%
        {DBLP:conf/kdd/HuangWZX21}
\bibfield{author}{\bibinfo{person}{Yanhua Huang}, \bibinfo{person}{Weikun Wang}, \bibinfo{person}{Lei Zhang}, {and} \bibinfo{person}{Ruiwen Xu}.} \bibinfo{year}{2021}\natexlab{}.
\newblock \showarticletitle{Sliding Spectrum Decomposition for Diversified Recommendation}. In \bibinfo{booktitle}{\emph{{KDD} '21: The 27th {ACM} {SIGKDD} Conference on Knowledge Discovery and Data Mining, Virtual Event, Singapore, August 14-18, 2021}}, \bibfield{editor}{\bibinfo{person}{Feida Zhu}, \bibinfo{person}{Beng~Chin Ooi}, {and} \bibinfo{person}{Chunyan Miao}} (Eds.). \bibinfo{publisher}{{ACM}}, \bibinfo{pages}{3041--3049}.
\newblock
\href{https://doi.org/10.1145/3447548.3467108}{doi:\nolinkurl{10.1145/3447548.3467108}}


\bibitem[Joachims et~al\mbox{.}(2007)]%
        {DBLP:journals/sigir/JoachimsLLZ07}
\bibfield{author}{\bibinfo{person}{Thorsten Joachims}, \bibinfo{person}{Hang Li}, \bibinfo{person}{Tie{-}Yan Liu}, {and} \bibinfo{person}{ChengXiang Zhai}.} \bibinfo{year}{2007}\natexlab{}.
\newblock \showarticletitle{Learning to rank for information retrieval {(LR4IR} 2007)}.
\newblock \bibinfo{journal}{\emph{{SIGIR} Forum}} \bibinfo{volume}{41}, \bibinfo{number}{2} (\bibinfo{year}{2007}), \bibinfo{pages}{58--62}.
\newblock
\href{https://doi.org/10.1145/1328964.1328974}{doi:\nolinkurl{10.1145/1328964.1328974}}


\bibitem[Lin et~al\mbox{.}(2019)]%
        {DBLP:conf/recsys/LinCPSXSZOJ19}
\bibfield{author}{\bibinfo{person}{Xiao Lin}, \bibinfo{person}{Hongjie Chen}, \bibinfo{person}{Changhua Pei}, \bibinfo{person}{Fei Sun}, \bibinfo{person}{Xuanji Xiao}, \bibinfo{person}{Hanxiao Sun}, \bibinfo{person}{Yongfeng Zhang}, \bibinfo{person}{Wenwu Ou}, {and} \bibinfo{person}{Peng Jiang}.} \bibinfo{year}{2019}\natexlab{}.
\newblock \showarticletitle{A pareto-efficient algorithm for multiple objective optimization in e-commerce recommendation}. In \bibinfo{booktitle}{\emph{Proceedings of the 13th {ACM} Conference on Recommender Systems, RecSys 2019, Copenhagen, Denmark, September 16-20, 2019}}, \bibfield{editor}{\bibinfo{person}{Toine Bogers}, \bibinfo{person}{Alan Said}, \bibinfo{person}{Peter Brusilovsky}, {and} \bibinfo{person}{Domonkos Tikk}} (Eds.). \bibinfo{publisher}{{ACM}}, \bibinfo{pages}{20--28}.
\newblock
\href{https://doi.org/10.1145/3298689.3346998}{doi:\nolinkurl{10.1145/3298689.3346998}}


\bibitem[Lin et~al\mbox{.}(2022)]%
        {DBLP:conf/kdd/LinWMZWJW22}
\bibfield{author}{\bibinfo{person}{Zihan Lin}, \bibinfo{person}{Hui Wang}, \bibinfo{person}{Jingshu Mao}, \bibinfo{person}{Wayne~Xin Zhao}, \bibinfo{person}{Cheng Wang}, \bibinfo{person}{Peng Jiang}, {and} \bibinfo{person}{Ji{-}Rong Wen}.} \bibinfo{year}{2022}\natexlab{}.
\newblock \showarticletitle{Feature-aware Diversified Re-ranking with Disentangled Representations for Relevant Recommendation}. In \bibinfo{booktitle}{\emph{{KDD} '22: The 28th {ACM} {SIGKDD} Conference on Knowledge Discovery and Data Mining, Washington, DC, USA, August 14 - 18, 2022}}, \bibfield{editor}{\bibinfo{person}{Aidong Zhang} {and} \bibinfo{person}{Huzefa Rangwala}} (Eds.). \bibinfo{publisher}{{ACM}}, \bibinfo{pages}{3327--3335}.
\newblock
\href{https://doi.org/10.1145/3534678.3539130}{doi:\nolinkurl{10.1145/3534678.3539130}}


\bibitem[Liu et~al\mbox{.}(2022)]%
        {DBLP:conf/ijcai/LiuXQS00ZT22}
\bibfield{author}{\bibinfo{person}{Weiwen Liu}, \bibinfo{person}{Yunjia Xi}, \bibinfo{person}{Jiarui Qin}, \bibinfo{person}{Fei Sun}, \bibinfo{person}{Bo Chen}, \bibinfo{person}{Weinan Zhang}, \bibinfo{person}{Rui Zhang}, {and} \bibinfo{person}{Ruiming Tang}.} \bibinfo{year}{2022}\natexlab{}.
\newblock \showarticletitle{Neural Re-ranking in Multi-stage Recommender Systems: {A} Review}. In \bibinfo{booktitle}{\emph{Proceedings of the Thirty-First International Joint Conference on Artificial Intelligence, {IJCAI} 2022, Vienna, Austria, 23-29 July 2022}}, \bibfield{editor}{\bibinfo{person}{Luc~De Raedt}} (Ed.). \bibinfo{publisher}{ijcai.org}, \bibinfo{pages}{5512--5520}.
\newblock
\href{https://doi.org/10.24963/IJCAI.2022/771}{doi:\nolinkurl{10.24963/IJCAI.2022/771}}


\bibitem[Miettinen(1998)]%
        {DBLP:books/daglib/0021267}
\bibfield{author}{\bibinfo{person}{Kaisa Miettinen}.} \bibinfo{year}{1998}\natexlab{}.
\newblock \bibinfo{booktitle}{\emph{Nonlinear multiobjective optimization}}. \bibinfo{series}{International series in operations research and management science}, Vol.~\bibinfo{volume}{12}.
\newblock \bibinfo{publisher}{Kluwer}.
\newblock
\showISBNx{978-0-7923-8278-2}


\bibitem[Nguyen and Salazar(2019)]%
        {DBLP:conf/iwslt/NguyenS19}
\bibfield{author}{\bibinfo{person}{Toan~Q. Nguyen} {and} \bibinfo{person}{Julian Salazar}.} \bibinfo{year}{2019}\natexlab{}.
\newblock \showarticletitle{Transformers without Tears: Improving the Normalization of Self-Attention}. In \bibinfo{booktitle}{\emph{Proceedings of the 16th International Conference on Spoken Language Translation, {IWSLT} 2019, Hong Kong, November 2-3, 2019}}, \bibfield{editor}{\bibinfo{person}{Jan Niehues}, \bibinfo{person}{Roldano Cattoni}, \bibinfo{person}{Sebastian St{\"{u}}ker}, \bibinfo{person}{Matteo Negri}, \bibinfo{person}{Marco Turchi}, \bibinfo{person}{Thanh{-}Le Ha}, \bibinfo{person}{Elizabeth Salesky}, \bibinfo{person}{Ramon Sanabria}, \bibinfo{person}{Lo{\"{\i}}c Barrault}, \bibinfo{person}{Lucia Specia}, {and} \bibinfo{person}{Marcello Federico}} (Eds.). \bibinfo{publisher}{Association for Computational Linguistics}.
\newblock
\urldef\tempurl%
\url{https://aclanthology.org/2019.iwslt-1.17}
\showURL{%
\tempurl}


\bibitem[Pei et~al\mbox{.}(2019)]%
        {DBLP:conf/recsys/PeiZZSLSWJGOP19}
\bibfield{author}{\bibinfo{person}{Changhua Pei}, \bibinfo{person}{Yi Zhang}, \bibinfo{person}{Yongfeng Zhang}, \bibinfo{person}{Fei Sun}, \bibinfo{person}{Xiao Lin}, \bibinfo{person}{Hanxiao Sun}, \bibinfo{person}{Jian Wu}, \bibinfo{person}{Peng Jiang}, \bibinfo{person}{Junfeng Ge}, \bibinfo{person}{Wenwu Ou}, {and} \bibinfo{person}{Dan Pei}.} \bibinfo{year}{2019}\natexlab{}.
\newblock \showarticletitle{Personalized re-ranking for recommendation}. In \bibinfo{booktitle}{\emph{Proceedings of the 13th {ACM} Conference on Recommender Systems, RecSys 2019, Copenhagen, Denmark, September 16-20, 2019}}, \bibfield{editor}{\bibinfo{person}{Toine Bogers}, \bibinfo{person}{Alan Said}, \bibinfo{person}{Peter Brusilovsky}, {and} \bibinfo{person}{Domonkos Tikk}} (Eds.). \bibinfo{publisher}{{ACM}}, \bibinfo{pages}{3--11}.
\newblock
\href{https://doi.org/10.1145/3298689.3347000}{doi:\nolinkurl{10.1145/3298689.3347000}}


\bibitem[Pi et~al\mbox{.}(2020)]%
        {DBLP:conf/cikm/PiZZWRFZG20}
\bibfield{author}{\bibinfo{person}{Qi Pi}, \bibinfo{person}{Guorui Zhou}, \bibinfo{person}{Yujing Zhang}, \bibinfo{person}{Zhe Wang}, \bibinfo{person}{Lejian Ren}, \bibinfo{person}{Ying Fan}, \bibinfo{person}{Xiaoqiang Zhu}, {and} \bibinfo{person}{Kun Gai}.} \bibinfo{year}{2020}\natexlab{}.
\newblock \showarticletitle{Search-based User Interest Modeling with Lifelong Sequential Behavior Data for Click-Through Rate Prediction}. In \bibinfo{booktitle}{\emph{{CIKM} '20: The 29th {ACM} International Conference on Information and Knowledge Management, Virtual Event, Ireland, October 19-23, 2020}}, \bibfield{editor}{\bibinfo{person}{Mathieu d'Aquin}, \bibinfo{person}{Stefan Dietze}, \bibinfo{person}{Claudia Hauff}, \bibinfo{person}{Edward Curry}, {and} \bibinfo{person}{Philippe Cudr{\'{e}}{-}Mauroux}} (Eds.). \bibinfo{publisher}{{ACM}}, \bibinfo{pages}{2685--2692}.
\newblock
\href{https://doi.org/10.1145/3340531.3412744}{doi:\nolinkurl{10.1145/3340531.3412744}}


\bibitem[Vaswani et~al\mbox{.}(2017)]%
        {DBLP:conf/nips/VaswaniSPUJGKP17}
\bibfield{author}{\bibinfo{person}{Ashish Vaswani}, \bibinfo{person}{Noam Shazeer}, \bibinfo{person}{Niki Parmar}, \bibinfo{person}{Jakob Uszkoreit}, \bibinfo{person}{Llion Jones}, \bibinfo{person}{Aidan~N. Gomez}, \bibinfo{person}{Lukasz Kaiser}, {and} \bibinfo{person}{Illia Polosukhin}.} \bibinfo{year}{2017}\natexlab{}.
\newblock \showarticletitle{Attention is All you Need}. In \bibinfo{booktitle}{\emph{Advances in Neural Information Processing Systems 30: Annual Conference on Neural Information Processing Systems 2017, December 4-9, 2017, Long Beach, CA, {USA}}}, \bibfield{editor}{\bibinfo{person}{Isabelle Guyon}, \bibinfo{person}{Ulrike von Luxburg}, \bibinfo{person}{Samy Bengio}, \bibinfo{person}{Hanna~M. Wallach}, \bibinfo{person}{Rob Fergus}, \bibinfo{person}{S.~V.~N. Vishwanathan}, {and} \bibinfo{person}{Roman Garnett}} (Eds.). \bibinfo{pages}{5998--6008}.
\newblock
\urldef\tempurl%
\url{https://proceedings.neurips.cc/paper/2017/hash/3f5ee243547dee91fbd053c1c4a845aa-Abstract.html}
\showURL{%
\tempurl}


\bibitem[Wang et~al\mbox{.}(2019)]%
        {DBLP:conf/acl/WangLXZLWC19}
\bibfield{author}{\bibinfo{person}{Qiang Wang}, \bibinfo{person}{Bei Li}, \bibinfo{person}{Tong Xiao}, \bibinfo{person}{Jingbo Zhu}, \bibinfo{person}{Changliang Li}, \bibinfo{person}{Derek~F. Wong}, {and} \bibinfo{person}{Lidia~S. Chao}.} \bibinfo{year}{2019}\natexlab{}.
\newblock \showarticletitle{Learning Deep Transformer Models for Machine Translation}. In \bibinfo{booktitle}{\emph{Proceedings of the 57th Conference of the Association for Computational Linguistics, {ACL} 2019, Florence, Italy, July 28- August 2, 2019, Volume 1: Long Papers}}, \bibfield{editor}{\bibinfo{person}{Anna Korhonen}, \bibinfo{person}{David~R. Traum}, {and} \bibinfo{person}{Llu{\'{\i}}s M{\`{a}}rquez}} (Eds.). \bibinfo{publisher}{Association for Computational Linguistics}, \bibinfo{pages}{1810--1822}.
\newblock
\href{https://doi.org/10.18653/V1/P19-1176}{doi:\nolinkurl{10.18653/V1/P19-1176}}


\bibitem[Wilhelm et~al\mbox{.}(2018)]%
        {DBLP:conf/cikm/WilhelmRBJCG18}
\bibfield{author}{\bibinfo{person}{Mark Wilhelm}, \bibinfo{person}{Ajith Ramanathan}, \bibinfo{person}{Alexander Bonomo}, \bibinfo{person}{Sagar Jain}, \bibinfo{person}{Ed~H. Chi}, {and} \bibinfo{person}{Jennifer Gillenwater}.} \bibinfo{year}{2018}\natexlab{}.
\newblock \showarticletitle{Practical Diversified Recommendations on YouTube with Determinantal Point Processes}. In \bibinfo{booktitle}{\emph{Proceedings of the 27th {ACM} International Conference on Information and Knowledge Management, {CIKM} 2018, Torino, Italy, October 22-26, 2018}}, \bibfield{editor}{\bibinfo{person}{Alfredo Cuzzocrea}, \bibinfo{person}{James Allan}, \bibinfo{person}{Norman~W. Paton}, \bibinfo{person}{Divesh Srivastava}, \bibinfo{person}{Rakesh Agrawal}, \bibinfo{person}{Andrei~Z. Broder}, \bibinfo{person}{Mohammed~J. Zaki}, \bibinfo{person}{K.~Sel{\c{c}}uk Candan}, \bibinfo{person}{Alexandros Labrinidis}, \bibinfo{person}{Assaf Schuster}, {and} \bibinfo{person}{Haixun Wang}} (Eds.). \bibinfo{publisher}{{ACM}}, \bibinfo{pages}{2165--2173}.
\newblock
\href{https://doi.org/10.1145/3269206.3272018}{doi:\nolinkurl{10.1145/3269206.3272018}}


\bibitem[Xia et~al\mbox{.}(2023)]%
        {DBLP:conf/www/XiaCHLL23}
\bibfield{author}{\bibinfo{person}{Yaxian Xia}, \bibinfo{person}{Yi Cao}, \bibinfo{person}{Sihao Hu}, \bibinfo{person}{Tong Liu}, {and} \bibinfo{person}{Lingling Lu}.} \bibinfo{year}{2023}\natexlab{}.
\newblock \showarticletitle{Deep Intention-Aware Network for Click-Through Rate Prediction}. In \bibinfo{booktitle}{\emph{Companion Proceedings of the {ACM} Web Conference 2023, {WWW} 2023, Austin, TX, USA, 30 April 2023 - 4 May 2023}}, \bibfield{editor}{\bibinfo{person}{Ying Ding}, \bibinfo{person}{Jie Tang}, \bibinfo{person}{Juan~F. Sequeda}, \bibinfo{person}{Lora Aroyo}, \bibinfo{person}{Carlos Castillo}, {and} \bibinfo{person}{Geert{-}Jan Houben}} (Eds.). \bibinfo{publisher}{{ACM}}, \bibinfo{pages}{533--537}.
\newblock
\href{https://doi.org/10.1145/3543873.3584661}{doi:\nolinkurl{10.1145/3543873.3584661}}


\bibitem[Xu et~al\mbox{.}(2023)]%
        {DBLP:conf/kdd/XuCWYSWHLZGH23}
\bibfield{author}{\bibinfo{person}{Yue Xu}, \bibinfo{person}{Hao Chen}, \bibinfo{person}{Zefan Wang}, \bibinfo{person}{Jianwen Yin}, \bibinfo{person}{Qijie Shen}, \bibinfo{person}{Dimin Wang}, \bibinfo{person}{Feiran Huang}, \bibinfo{person}{Lixiang Lai}, \bibinfo{person}{Tao Zhuang}, \bibinfo{person}{Junfeng Ge}, {and} \bibinfo{person}{Xia Hu}.} \bibinfo{year}{2023}\natexlab{}.
\newblock \showarticletitle{Multi-factor Sequential Re-ranking with Perception-Aware Diversification}. In \bibinfo{booktitle}{\emph{Proceedings of the 29th {ACM} {SIGKDD} Conference on Knowledge Discovery and Data Mining, {KDD} 2023, Long Beach, CA, USA, August 6-10, 2023}}, \bibfield{editor}{\bibinfo{person}{Ambuj~K. Singh}, \bibinfo{person}{Yizhou Sun}, \bibinfo{person}{Leman Akoglu}, \bibinfo{person}{Dimitrios Gunopulos}, \bibinfo{person}{Xifeng Yan}, \bibinfo{person}{Ravi Kumar}, \bibinfo{person}{Fatma Ozcan}, {and} \bibinfo{person}{Jieping Ye}} (Eds.). \bibinfo{publisher}{{ACM}}, \bibinfo{pages}{5327--5337}.
\newblock
\href{https://doi.org/10.1145/3580305.3599869}{doi:\nolinkurl{10.1145/3580305.3599869}}


\bibitem[Yang et~al\mbox{.}(2005)]%
        {DBLP:conf/airs/YangJL05}
\bibfield{author}{\bibinfo{person}{Lingpeng Yang}, \bibinfo{person}{Dong{-}Hong Ji}, {and} \bibinfo{person}{Mun{-}Kew Leong}.} \bibinfo{year}{2005}\natexlab{}.
\newblock \showarticletitle{Chinese Document Re-ranking Based on Term Distribution and Maximal Marginal Relevance}. In \bibinfo{booktitle}{\emph{Information Retrieval Technology, Second Asia Information Retrieval Symposium, {AIRS} 2005, Jeju Island, Korea, October 13-15, 2005, Proceedings}} \emph{(\bibinfo{series}{Lecture Notes in Computer Science}, Vol.~\bibinfo{volume}{3689})}, \bibfield{editor}{\bibinfo{person}{Gary~Geunbae Lee}, \bibinfo{person}{Akio Yamada}, \bibinfo{person}{Helen Meng}, {and} \bibinfo{person}{Sung{-}Hyon Myaeng}} (Eds.). \bibinfo{publisher}{Springer}, \bibinfo{pages}{299--311}.
\newblock
\href{https://doi.org/10.1007/11562382\_23}{doi:\nolinkurl{10.1007/11562382\_23}}


\bibitem[Yang et~al\mbox{.}(2021)]%
        {DBLP:conf/cvpr/YangZQ021}
\bibfield{author}{\bibinfo{person}{Xu Yang}, \bibinfo{person}{Hanwang Zhang}, \bibinfo{person}{Guojun Qi}, {and} \bibinfo{person}{Jianfei Cai}.} \bibinfo{year}{2021}\natexlab{}.
\newblock \showarticletitle{Causal Attention for Vision-Language Tasks}. In \bibinfo{booktitle}{\emph{{IEEE} Conference on Computer Vision and Pattern Recognition, {CVPR} 2021, virtual, June 19-25, 2021}}. \bibinfo{publisher}{Computer Vision Foundation / {IEEE}}, \bibinfo{pages}{9847--9857}.
\newblock
\href{https://doi.org/10.1109/CVPR46437.2021.00972}{doi:\nolinkurl{10.1109/CVPR46437.2021.00972}}


\bibitem[Zeng et~al\mbox{.}(2021)]%
        {DBLP:conf/aaai/ZengYHNLZM21}
\bibfield{author}{\bibinfo{person}{Anxiang Zeng}, \bibinfo{person}{Han Yu}, \bibinfo{person}{Hua{-}Lin He}, \bibinfo{person}{Yabo Ni}, \bibinfo{person}{Yongliang Li}, \bibinfo{person}{Jingren Zhou}, {and} \bibinfo{person}{Chunyan Miao}.} \bibinfo{year}{2021}\natexlab{}.
\newblock \showarticletitle{Enhancing E-commerce Recommender System Adaptability with Online Deep Controllable Learning-To-Rank}. In \bibinfo{booktitle}{\emph{Thirty-Fifth {AAAI} Conference on Artificial Intelligence, {AAAI} 2021, Thirty-Third Conference on Innovative Applications of Artificial Intelligence, {IAAI} 2021, The Eleventh Symposium on Educational Advances in Artificial Intelligence, {EAAI} 2021, Virtual Event, February 2-9, 2021}}. \bibinfo{publisher}{{AAAI} Press}, \bibinfo{pages}{15214--15222}.
\newblock
\href{https://doi.org/10.1609/AAAI.V35I17.17785}{doi:\nolinkurl{10.1609/AAAI.V35I17.17785}}


\bibitem[Zhai et~al\mbox{.}(2024)]%
        {DBLP:conf/icml/ZhaiLLWLCGGGHLS24}
\bibfield{author}{\bibinfo{person}{Jiaqi Zhai}, \bibinfo{person}{Lucy Liao}, \bibinfo{person}{Xing Liu}, \bibinfo{person}{Yueming Wang}, \bibinfo{person}{Rui Li}, \bibinfo{person}{Xuan Cao}, \bibinfo{person}{Leon Gao}, \bibinfo{person}{Zhaojie Gong}, \bibinfo{person}{Fangda Gu}, \bibinfo{person}{Jiayuan He}, \bibinfo{person}{Yinghai Lu}, {and} \bibinfo{person}{Yu Shi}.} \bibinfo{year}{2024}\natexlab{}.
\newblock \showarticletitle{Actions Speak Louder than Words: Trillion-Parameter Sequential Transducers for Generative Recommendations}. In \bibinfo{booktitle}{\emph{Forty-first International Conference on Machine Learning, {ICML} 2024, Vienna, Austria, July 21-27, 2024}}. \bibinfo{publisher}{OpenReview.net}.
\newblock
\urldef\tempurl%
\url{https://openreview.net/forum?id=xye7iNsgXn}
\showURL{%
\tempurl}


\bibitem[Zhang and Hurley(2008)]%
        {DBLP:conf/recsys/ZhangH08}
\bibfield{author}{\bibinfo{person}{Mi Zhang} {and} \bibinfo{person}{Neil Hurley}.} \bibinfo{year}{2008}\natexlab{}.
\newblock \showarticletitle{Avoiding monotony: improving the diversity of recommendation lists}. In \bibinfo{booktitle}{\emph{Proceedings of the 2008 {ACM} Conference on Recommender Systems, RecSys 2008, Lausanne, Switzerland, October 23-25, 2008}}, \bibfield{editor}{\bibinfo{person}{Pearl Pu}, \bibinfo{person}{Derek~G. Bridge}, \bibinfo{person}{Bamshad Mobasher}, {and} \bibinfo{person}{Francesco Ricci}} (Eds.). \bibinfo{publisher}{{ACM}}, \bibinfo{pages}{123--130}.
\newblock
\href{https://doi.org/10.1145/1454008.1454030}{doi:\nolinkurl{10.1145/1454008.1454030}}


\bibitem[Zhou et~al\mbox{.}(2018)]%
        {DBLP:conf/kdd/ZhouZSFZMYJLG18}
\bibfield{author}{\bibinfo{person}{Guorui Zhou}, \bibinfo{person}{Xiaoqiang Zhu}, \bibinfo{person}{Chengru Song}, \bibinfo{person}{Ying Fan}, \bibinfo{person}{Han Zhu}, \bibinfo{person}{Xiao Ma}, \bibinfo{person}{Yanghui Yan}, \bibinfo{person}{Junqi Jin}, \bibinfo{person}{Han Li}, {and} \bibinfo{person}{Kun Gai}.} \bibinfo{year}{2018}\natexlab{}.
\newblock \showarticletitle{Deep Interest Network for Click-Through Rate Prediction}. In \bibinfo{booktitle}{\emph{Proceedings of the 24th {ACM} {SIGKDD} International Conference on Knowledge Discovery {\&} Data Mining, {KDD} 2018, London, UK, August 19-23, 2018}}, \bibfield{editor}{\bibinfo{person}{Yike Guo} {and} \bibinfo{person}{Faisal Farooq}} (Eds.). \bibinfo{publisher}{{ACM}}, \bibinfo{pages}{1059--1068}.
\newblock
\href{https://doi.org/10.1145/3219819.3219823}{doi:\nolinkurl{10.1145/3219819.3219823}}


\bibitem[Zhou et~al\mbox{.}(2022)]%
        {DBLP:conf/www/ZhouYZW22}
\bibfield{author}{\bibinfo{person}{Kun Zhou}, \bibinfo{person}{Hui Yu}, \bibinfo{person}{Wayne~Xin Zhao}, {and} \bibinfo{person}{Ji{-}Rong Wen}.} \bibinfo{year}{2022}\natexlab{}.
\newblock \showarticletitle{Filter-enhanced {MLP} is All You Need for Sequential Recommendation}. In \bibinfo{booktitle}{\emph{{WWW} '22: The {ACM} Web Conference 2022, Virtual Event, Lyon, France, April 25 - 29, 2022}}, \bibfield{editor}{\bibinfo{person}{Fr{\'{e}}d{\'{e}}rique Laforest}, \bibinfo{person}{Rapha{\"{e}}l Troncy}, \bibinfo{person}{Elena Simperl}, \bibinfo{person}{Deepak Agarwal}, \bibinfo{person}{Aristides Gionis}, \bibinfo{person}{Ivan Herman}, {and} \bibinfo{person}{Lionel M{\'{e}}dini}} (Eds.). \bibinfo{publisher}{{ACM}}, \bibinfo{pages}{2388--2399}.
\newblock
\href{https://doi.org/10.1145/3485447.3512111}{doi:\nolinkurl{10.1145/3485447.3512111}}


\end{thebibliography}

\section*{Main Author Bio}
\textbf{Yue Meng} is a researcher in the Department of Search and
Recommendation at Taobao \& Tmall Group of Alibaba. He received his master degree from Peking University. His research focuses on recommendation system.

\noindent \textbf{Cheng Guo} is a researcher in the Department of Search and
Recommendation at Taobao \& Tmall Group of Alibaba. He received his master degree from Tsinghua University, where he was part of THUIR(Information Retrieval Lab at Tsinghua University). His research focuses on information retrieval.

\noindent \textbf{Yi Cao} is the leader of Marketing Algorithm in the Department of Search and Recommendation at Taobao \& Tmall Group of Alibaba. He received his master degree from Zhejiang University. His research focuses on recommendation system.

\end{document}